# A Predictive Model for Student Performance in Classrooms Using Student Interactions With an eTextbook

*Ahmed Abd Elrahman, Taysir Hassan A Soliman, Ahmed I. Taloba, and Mohammed F. Farghally*

Information System Department, Faculty of Computers and Information, Assiut University, Egypt

ahmedabdo@aun.edu.eg, taysirhs@aun.edu.eg, Taloba@aun.edu.eg, mfseddik@aun.edu.eg

**Abstract**

With the rise of online eTextbooks and Massive Open Online Courses (MOOCs), a huge amount of data has been collected related to students' learning. With the careful analysis of this data, educators can gain useful insights into the performance of their students and their behavior in learning a particular topic. This paper proposes a new model for predicting student performance based on an analysis of how students interact with an interactive online eTextbook. By being able to predict students' performance early in the course, educators can easily identify students at risk and provide a suitable intervention. We considered two main issues the prediction of good/bad performance and the prediction of the final exam grade. To build the proposed model, we evaluated the most popular classification and regression algorithms on data from a data structures and algorithms course (CS2) offered in a large public research university. Random Forest Regression and Multiple Linear Regression have been applied in Regression. While Logistic Regression, decision tree, Random Forest Classifier, K Nearest Neighbors, and Support Vector Machine have been applied in classification. Based on the findings of the experiments, the algorithm with the best result overall in classification for this data was Random Forest Classifier with an accuracy equal to 91.7% while in the regression it was Random Forest Regression with an $R^2$ equal to 0.977.

## 1. Introduction

With the rapid spread of COVID-19 over the world.it has had a significant impact on human life, both current era and in the coming years. It has become necessary to find alternative learning ways rather than traditional classroom learning. Online eTextbooks and Massive Open Online Courses (MOOCs) have become the best alternative to learning ways. There is more potential in Technology-enhanced environments than traditional classroom learning.

When teaching a particular course via a Technology-enhanced environment, the interaction between students and instructors may be limited and there may be no face-to-face interaction. Therefore, it may be difficult for educators to know the performance of their students especially low-risk students with learning disabilities, and who may leave the course. So, the prediction of students' performance at an early stage in the course might be helpful in detecting those students with a high risk of failing the course. an early prediction maybe serve as an active tool for changing educators' practices and issuing an awareness to assist students to Back on the right track. In the discipline of learning analytics, the early prediction of student performance is a significant task. We think that it will improve academic retention and performance.

An early prophecy enables students to take the required stages to evade poor performance and enhance their own test achievements ahead of time. This early prediction not only alerts students to their poor performance but also gives them plenty of bases to maximize their academic performance [1].

Our study focused on a CS2 course. This course was taught at a large public research institution using the OpenDSA eTextbook infrastructure [5, 6]. OpenDSA is an infrastructure to create eTextbooks including interactive visualizations and exercises with automated feedback. OpenDSA gathers log data for all user interactions that happen on an OpenDSA module. Computer science students tend to have high drop-out rates as compared to other disciplines.

According to an article published in Irish Times [2] After their first year of college, in all institutes of technology, about one-third of computer science students drop out." According to paper [3] 30-50% of students at the Helsinki University of Technology did dropping out from the programming in Java course (CS1).

According to this article [4], there are various reasons why students drop out of courses, such as expensive tuition, not being prepared scholastically, being unhappy with the college, dissuading surroundings, selecting the wrong topic, academic inadequacy, and conflict with work and family responsibilities. There may be some students who are affected by one or more of these reasons, but it is clear that those students require extra effort to thrive in their courses. Instructors should be able to identify these students early in the course to assist them in improving their performance.

In this paper [7], A questionnaire was sent to computer science educators asking them to list topics that they believe are important for students to learn as well as topics that are difficult to learn (for students) and difficult to teach (for instructors). Based on the findings of this questionnaire, the educators found that the five most topics which hard to learn are pointers, recursion, polymorphism, memory allocation, and parameter passing. While they found that Recursion, pointers, error handling techniques, and polymorphism were the most difficult topics to teach. Most of these topics were taught in the CS2 course.

According to the conclusions of the questionnaire from the paper [7], many of the topics in the CS2 course are difficult to learn and teach. Many students may find difficulty in learning and understanding these topics, and they may not report their difficulties in learning these topics to their educators. Due to the difficulty of most topics of the CS2 course and based on the paper [3] which mentioned that the difficulty of the course was one of the compelling reasons that made students drop out from the CS1 course, many students maybe resort to dropping out of this course. Therefore, a method must be devised to aid educators in learning about each student's performance in this course, and it is preferable that this knowledge become at an early stage of the course.

Our study intends to build a model for predicting student performance based on their interactions with an eTextbook. This predictive approach has the advantage that it can be used at the beginning of the semester, if necessary, by instructors to communicate their concerns to students when there are signs of hazard. As a result, those students graduate on time and without having to repeat a semester, and they are well-prepared to succeed in their subsequent studies. This early prediction may be helpful to instructors to know early feedback to each student in the course, particularly low-risk, students with learning disabilities who need special attention, and maybe consider dropping out of this course. This feedback may assist instructors in providing appropriate warnings or advising to these students providing more attention to them to enhance their performance and trying to keep students from dropping out of the course.it may allow educators to help these students, and they may succeed in helping students to increase their academic performance and in reducing the falling ratio.

One of the advantages of the study is, In order to identify students' performance, our predictive approach does not require educators to give quizzes or midterm exams. Because the characteristics gathered in this study and those used to build the model focus on the extent to which students engage with the eTextbook and the amount of time each student spends with the eTextbook.

The structure of this paper is as follows: In Section2, previous works relating to students' performance is reviewed. Section 3 presented the data set description. Section 4 discusses the outcomes of various data mining approaches. Section 5 presented the Conclusion. Section 6 presented the future work.

## 2. Related Work

In [8], a study was conducted to predict students' grades in their work and their results (pass/fail). To predict the students' results, a classification model was utilized, while a regression model was used to predict the grades. For classification, decision trees and SVM were used, and for regression analysis, SVM Random Forest and AdaBoost.R2 were used. In this study, the classification model was shown to be capable of extracting useful patterns, while regression methods failed to prevail over a simple baseline.

A study has been conducted depending on the performance of students by choosing students from various institutes of Dr.R.M.L. Awadh University, Faizabad, India by using

Bayes Classification on Category, Language, and Background Qualification. The goal of this study is to determine if incoming students will perform or not, as well as to determine which students who require special attention in order to lower the falling rate [9].

According to a study published in [10], a decision tree model was used to determine the final mark for students enrolled in a C++ course at Yarmouk University in Jordan. Three classification methods were used: ID3, C4.5, and Naive Bayesian. The results showed that the decision tree model outperformed the other models in terms of prediction.

A case study has been conducted in [11] the students' data was used to analyze their learning in order to forecast their outcomes and warn students who could be in danger before their final exams.

In [12], a machine learning method have been used to enhance the prognosis results of academic achievements in real-world case studies. Three methods have been used to resolve the problem of class imbalance all of which gave positive results. After balancing the datasets, both cost-sensitive and insensitive learning algorithms were used, including SVM for small datasets and a Decision tree for large datasets.

In [13], a design for student careers is developed. This paper described various methods based on sequential and clustering pattern algorithms to propose solutions for enhancing student performance and exam schedule.

According to a study published in [14], data mining techniques have been used to investigate students' intellectual performance. The primary objective of this study is to assess the student's performance across a variety of measurement categories.

In [15], classification-based techniques have been utilized to predict students who are slow learners. Five classification techniques have been used: Multilayer Perceptron, Naive Bayes, SMO, J48, and REP Tree to analyze and test the output dataset. According to the findings of this study, it was found that Multilayer Perception outperforms all other classifiers.

To improve students' performance at an early stage, a study is made in [16], which may predict student's performance at an early stage and provide early warning to these students. Three well-known single classification algorithms, C4.5, CART, and LGR have been used to design this system.

In [17], a model is developed for predicting student achievement based on students' personal, pre-university, and university functional properties, with the neural network obtaining the highest accuracy, followed by the decision tree classifier and the kNN model.

The authors of the paper [18] presented a study in which they applied data mining techniques on educational data by analyzing students' academic performance. They used the decision tree approach to identify dropouts and students who require further assistance, making it easier for instructors to issue warnings or advice.

The next sections will describe the OpenDSA system in detail, the Data set used in our work and its descriptions, the data mining algorithms, and their results.

## 3. MATERIALS and Data Set Description

During spring 2020, OpenDSA was used as the main eTextbook to teach CS2 course in a large public research institution. A module in OpenDSA represents a single topic or part of a typical lecture, such as a single sorting algorithm and it is considered the most elementary functional unit for OpenDSA materials [19].

Every module is a full unit of instruction that usually contains algorithm visualizations (AVs), interactive assessment activities with automated feedback, and textbook quality text. Modules are organized into chapters in the same way that traditional paper books are organized. One of the most important OpenDSA exercises is "Algorithm simulations". These exercises ask the students to handle a data structure in order to demonstrate how an algorithm works, such as clicking on proper nodes in a tree or clicking to swap elements in an array. The JavaScript Algorithm Visualization (JSAV) [20] library is used to build this type of exercise.

OpenDSA contains multiple types of exercises and events. The next section describes exercise types and events.

### 3.1 OpenDSA events and exercises

In this work, about 200 different types of events were recorded by the OpenDSA system. The following is a summary of these events.

- Interactions with sign-up and sign-in
- Interactions with Static Content (such that when a module page is loaded by a student, when a student uses the navigation menu to navigate to another page)
- Reactive Activities Interactions (when a student needs to go ahead a slideshow)
- Interactions between Assessment Activities (such that when an exercise is loaded or when the answer is submitted by students, etc.)
- Interactions in the grade book (when a student need to know his score so he loads the gradebook page)

There is a timestamp for each event. To determine the relationship between the use of OpenDSA and student performance, log data and performance data (student grades written "etest") was utilized. The visualization was frequently used in the classroom as a lecture assist by the instructor, who utilized OpenDSA as the primary course material.

OpenDSA provides three types of exercises: proficiency exercises, simple questions, and programming exercises. For Proficiency exercises, this type of exercise is an algorithm simulation exercise. It requires students to simulate the behavior of a given algorithm in order to ensure that they understand how it works. This type of exercise was pioneered in the TRAKLA2 system [21].

Simple questions are made up of a variety of OpenDSA system question categories, including true/false, multiple-choice, and fill-in-the-blank, and they don't typically take a long time to finish. To store and present the simple exercises, OpenDSA used the Khan Academy framework [22].

All exercises are assessed automatically and comments were provided to the students. Students can work on exercises more than once until they obtain credits.

Slideshows is a type of interactive content provided by OpenDSA. Slideshows show a sequence of stages that animate the behavior of an algorithm. Slideshows were created using JSAV: the JavaScript Algorithm Visualization Library [20]. Figure 1 shows an example of an OpenDSA slideshow.

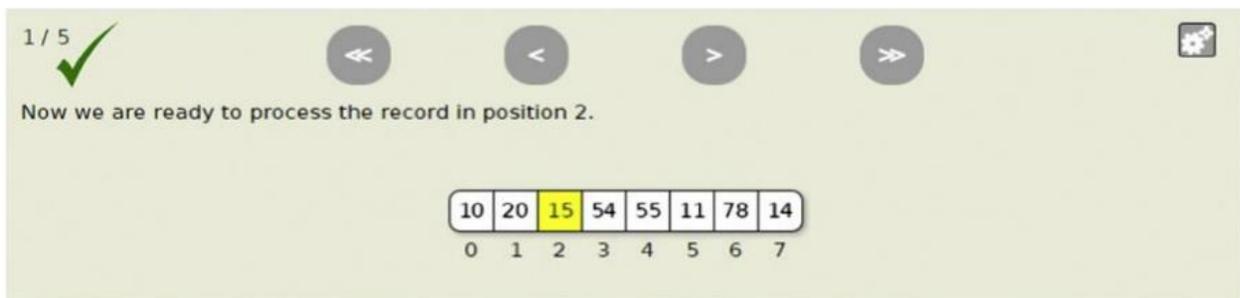

**Figure 1.** An example of an OpenDSA slideshow

Standard controls, as illustrated in Figure 1, allow the user to advance the slideshow by one slide, back up one slide, go back to the beginning, or leap to the finish of the slideshow.

As shown in Figure 2 and Figure 3, for each Proficiency exercise, the student can ask for a model answer or reset the exercise. For each Simple exercise, the student can ask for a hint or check his or her answer for this question. A student can attempt to complete each exercise category many times. So, we took the maximum number of attempts for each question.

**Figure 2**. Example of an OpenDSA simple question.

**Figure 3**. Example of JSAV proficiency exercises for Heap sort.

In this work, for Simple exercises, the student can ask for a hint more than one time to the same question. We have taken the maximum number of hints for each simple exercise. For the slideshows, each student can view the same slide more than one time. So, we have stored the total number of visits in the field called (total visit), but the unique number of slides that students viewed is stored in a field called (slide). We didn't include Programming Exercises in this study.

**3.2. Dataset Description**

We collected and analyzed data from 200 students. We gathered the following attributes for each student. Table 1 describes a summary of them.

**Table 1**. Date set Description**.**

| Attribute No | Attribute Name | Attribute Description |
|---|---|---|
| 1 | PE_total_time | Total time in seconds which a student was spent in solving or trying to solve proficiency exercises. |
| 2 | PE_total_attempts | The total number of attempts a student made to complete proficiency exercises. |
| 3 | PE_reset | How many times did a student reset their proficiency exercises? |
| 4 | PE_model | For proficiency exercises, the total number of times a student showed the model answer. |
| 5 | PE_exercise | The total number of proficiency exercises did each student solve |
| 6 | SS_total_time | The Total time in seconds which a student was taken in viewing slideshows. |
| 7 | SS_total_visit | A student's total number of slideshows seen; a student can watch the same slideshow many times. |
| 8 | slide | The total number of unique slideshows which a student has viewed. |
| 9 | Interaction | The total number of interactions that a student did. |
| 10 | Total_time | The total time in seconds has been spent by the student in dealing with the whole eTextbook. |
| 11 | Total_attempts | A total number of times did a student try to solve a Simple exercise |
| 12 | Total_hints | The total number of hints did a student use during solving Simple exercises. |
| 13 | gaming | The total number of pages reloads for each student |
| 14 | exercise | The Total number of exercises correctly completed. |
| 15 | etest | The final exam degree. |

## 4. Data Mining Techniques and Results

Many application domains have used data mining techniques, including banking, fraud detection, and telecommunications [23]. Data mining approaches have recently been utilized to improve and evaluate higher education problems.

In learning environments, the capacity to determine a student's performance is critical. The utilization of Data Mining is a very promising approach for achieving this goal [24]. Data mining techniques are applied to large amounts of data to uncover hidden patterns and relationships that aid decision-making.

The goal of our study is to predict (before the final exam) the performance (good/bad) and the final exam degree of students in one of the undergraduate data structures courses at a large public research institution at an early stage. This early prediction will aid in the identification of low-risk students at any point during the course, not only at the end. This early prediction aims to help these students to overcome the challenges they face in the learning process. This early prediction helps students work on their weaknesses to improve their performance and obtain good grades in their final exam. Furthermore, the findings will aid teachers in revising their instructional practices to improve student learning and prevent them from dropping out of this course. To achieve our goal, we address two problems the first is the prediction of student performance and the second is the prediction of students' final score in the final exam. We used two data mining approaches to achieve our goal. The first one is regression analysis which has the purpose of predicting the final degree for the exam. While the second one is Classification which has the purpose of predicting the performance for every student whether the performance will be good or bad. Figure 4 shows the steps of different date mining approaches.

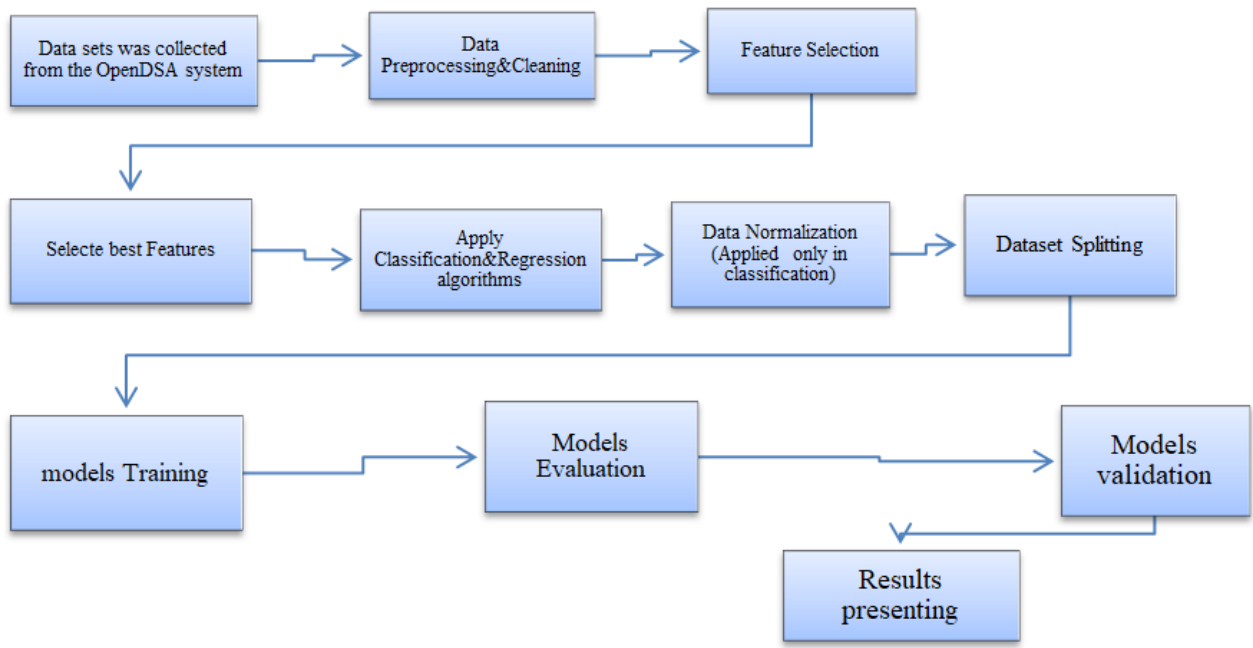

**Figure 4**. Steps of data mining approaches.

**Data Preprocessing**

Prior to implementing a data mining technique, data preprocessing turns the original data into a form that can be utilized by a specific data mining algorithm. Data preprocessing entails a variety of tasks. Data cleaning, feature selection, and data transformation are all steps in the data preprocessing process [25].

**Data Cleaning**

It is regarded as one of the most crucial data preprocessing steps. Data cleansing, often known as scrubbing, is the process of discovering and removing errors and inconsistencies from data in order to enhance data quality. The purpose of data cleaning is to clean up the data by removing irrelevant items and missing values. After removing the missing records from the dataset, the dataset was reduced to 194 records. We have applied data cleaning in both regression and classification algorithms.

## 4.1. Regression

Regression analysis is a group of statistical processes for appreciating the relationships between a dependent variable (often called the 'outcome') and one or more independent variables (often called features or predictors). As a result of the regression, the predictor is a continuous variable. In our study, the attribute "etest" is the dependent variable while the other attributes are the predictors. Linear regression is the most prevalent regressor in educational data mining [26]. However, regression trees are also very common. There are fewer regressors such as support vector machines and neural networks utilized in educational data mining than in other disciplines [26]. This is thought that more conservative algorithms are more successful in educating domains because of the high levels of noise and various explanatory elements [26]. Multiple linear regression and random forest regression have been utilized as regression techniques in our study.

### 4.1.1 Feature Selection

The main task in a data preprocessing area is feature selection. While eliminating redundant and irrelevant data is the goal of feature selection, selecting an adequate subset of features that can proficiently describe input data minimizes the feature space's size [27]. As a result, this process can have a considerable impact on the learning algorithm's efficiency.

Wrapper-based and filter-based techniques are the two types of feature selection methods used in supervised learning. Filter-based approaches are used to evaluate the relationship between input variables and the target variable, and the scores from these evaluations are used to select (filter) which input variables will be included and will be used in the model.

For filter feature selection methods, there are many methods such that Correlation Feature Selection and Mutual information feature selection. About Correlation Feature Selection, A measure of how two variables change together is called correlation. For numeric predictors, the sample correlation statistic is the classical approach used to quantify each relationship with the outcome. To determine the correlation between two random variables, there are two broad categories to consider. The first, which is based on a linear correlation, the second is based on an information theory. Among these two measures, the linear coefficient of correlation is the most familiar. In general, linear correlation scores range from -1 to 1, with 0 representing no relationship between the two characteristics being analyzed. We're generally looking for a positive score for feature selection, the higher the positive value, the stronger the relationship and the more likely the feature will be used in modeling. Standard literature states that the linear correlation coefficient "r" for two variables (A, B) is given by:

$$r = \frac{\sum(AB) - n\overline{A}\overline{B}}{(n-1)\sigma_A \sigma_B} \tag{1}$$

Where n is the number of A and B values, $\overline{A}$ and $\overline{B}$ are the means of A and B variables, $\sigma_A$ is the standard of A variable, $\sigma_B$ is the standard deviation of B variable, and $\Sigma$ (AB) is the sum of the AB cross-product.

A correlation feature selection approach was applied in this study in order to determine which features were most important while building a model that would predict students' final exam degree. Correlation Feature Selection scores are shown in Figure 4, after all, features were scored using this method. It has been determined that the first eleven features have a high influence on the outcome of the regression algorithms. As a result, those features were chosen, while others were not.

As shown in Figure 5, the PE_exercise feature got the highest score, and then followed by Total_attempts, PE_total_attempts, gaming, Total_hints, PE_total_time, total_time, SS_total_visit, interaction, slide, PE_model, PE_reset, SS_total_time, and exercise. We have selected the first eleven features which have the highest score while other ones are excluded. We excluded exercise, SS_total_time, and PE_reset features.

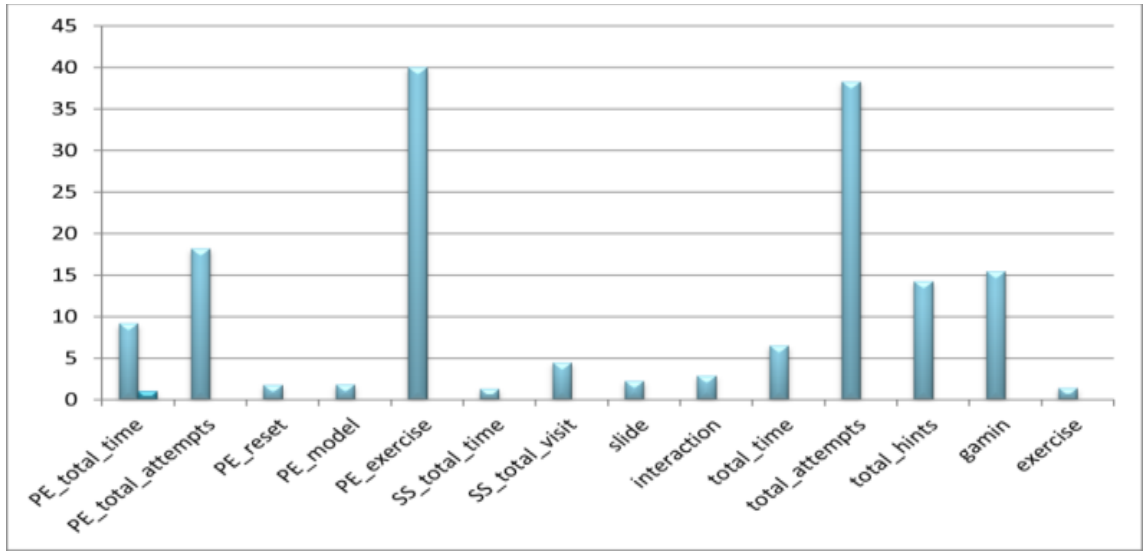

**Figure 5**. The score for each feature after applying the correlation feature selection method.

### 4.1.2 Methodology and Results

The regression techniques were applied after the feature selection process. In the next section, we will review the two regression algorithms used in this study with their Accuracy assessment ended with the Results.

### A. Multiple Linear Regression (MLR)

Multiple linear regression is a technique for understanding the relationship between two or more independent variables (or predictors) and one continuous dependent variable. In our study, the dependent (or outcome) variable is the 'etest' variable while the other features are the independent variables. Linear regression models generally have the form:

$$y = a + \sum_{i=1}^{n} b_i * x_i + €  \quad (2)$$

In this case, "y" is the dependent variable (etest), "x" is the predictors, a is y-intercept (constant term), € is the model's error term, and "n" is the number of independent variables. Based on the independent variables that are available, the regression equation was used to forecast students' final grades.

### B. Random Forest Regression (RFR)

RFR is considered one of the ensemble machine learning approachs. RFR predicts an outcome from a set of predictors by creating multiple Decision Trees (DTs) and aggregating their results. By utilizing a unique bootstrap sample of the training data, each tree in a forest is created independently. Instead of using the best split among all predictors (as in bagging and bootstrapping [28]) for node splitting, RF chooses the best split from a randomly selected subset of predictors. The addition of this randomization reduces the association between trees in the forest so this operation will increase accuracy [29].

**C. Accuracy assessment**

This section presents the Accuracy assessment methods used to determine the efficiency for the two regression algorithms used here. We utilized three separate metrics to assess the quality of the regression algorithms: Mean absolute percentage error, root mean squared error, and coefficient of determination.

**1-Coefficient of Determination or R Squared ($R^2$)**

($R^2$ or r-squared) is a statistical measure in a regression model that predicts the proportion of the difference in the dependent variable that can be described by the independent variable. In other words, the coefficient of determination shows how well the data fit the model (goodness of fit). The most typical interpretation of $R^2$ is how well the regression model fits the observed data. A higher coefficient indicates a better model fit for the model. A model with an $R^2$ score of 1 is perfect, whereas a score of 0 indicates that it will perform badly on an unknown dataset. This also means that the closer the r squared score is to 1, the better the model has been trained. The following formula is used to compute $R^2$:

$$R^2 = 1 - \frac{SSE}{SST} \quad (3)$$

Where, SSE is called the Total sum of squares, and

$$SSE = \sum_{i=1}^{N} (y_{actual_i} - y_{mean})^2 \quad (4)$$

Where $y_{actual_i}$ is the original or observed y-value, $y_{mean}$ is the mean of y-value.

-SST is called the sum of squares due to regression, and

$$SST = \sum_{i=1}^{N} (y_{predicted_i} - y_{mean})^2 \quad (5)$$

Where $y_{predicted}$ is the y-value of regression and $y_{mean}$ is the mean of y-value.

The variation in the observed data is measured using the SSE. We can determine how well the model represents the data that was utilized in the model by using the SST

**2. Mean Absolute Percentage Error (MAPE)**

MAPE is the mean or average of the absolute percentage errors of forecasts. The difference between the actual or observed value and the predicted value is called an error. And it is defined as a measure of predictive accuracy of a prediction technique in statistics. The better the forecast, the less the MAPE. The following formula shows how to compute the MAPE value.

$$\text{MAPE} = \frac{100}{N} \sum_{i=1}^{N} \left| \frac{A_i - P_i}{A_i} \right| \quad (6)$$

Where N is the number of items $A_i$ is the actual value and $P_i$ is the predicted value.

### 3. Root Mean Squared Error (RMSE)

RMSE is a repeatedly utilized measure of the differences between values that the model predicted and the values observed. RMSD represents the square root of the discrepancies between anticipated and observed values, or the quadratic mean of these differences. We can compute RMSE using the following formula:

$$\text{RMSE} = \sqrt{\frac{\sum_{i=1}^{N} (\text{predicted}_i - \text{Actual}_i)^2}{N}} \quad (7)$$

Where N = number of items, $\text{Actual}_i$ is the original or observed y-value, $\text{predicted}_i$ is the predicted value.

### D. Regression algorithms Results

According to the partition of data set in paper [30]. The data set was separated into ratios of 80% to 20%, having trained data about 80% and testing data about 20%. Random student data was also utilized to predict grades to evaluate the model.

**Table 2** Shows the results which we obtained after applying MLR and RFR to our dataset using selected features and without using them.

Table 2 shows that RFR has the highest R2 value, as well as the lowest RMSE and MAPE values, followed by MLR. The R2 for the two algorithms is very good when using all features or only the eleven selected features but both of them give better results when we use only the eleven selected features. As we said earlier in this paper, these results proved that feature selection improve the algorithm's efficiency.

**Table 2**. Regression models results.

| Algorithm | $R^2$ | RMSE | MAPE |
|---|---|---|---|
| MLR with All features | 0.918 | 8.937 | 5.01 |
| MLR with eleven selected features | 0.922 | 8.759 | 4.89 |
| RFR with All features | 0.977 | 4.821 | 2.47 |
| RFR with eleven selected features | 0.976 | 4.799 | 2.27 |

## 4.2. Classification

Classification is one of the most often utilized and studied data mining techniques. Classification is a technique for predicting the class or category of a data object based on previously learned classes from a training dataset with known classes. We divided the students into two groups based on their final exam's degree, which are:

Label: "**good**" for grades above 65% from the final exam degree.

Label: "**bad**" for grades less than or equal to 65% from the final exam degree. As sown in Table 3

**Table 3.** Selecting 2 class label according to student's grades

| class | Grade | Number of students | percentage |
|---|---|---|---|
| good | >65% | 181 | 93% |
| bad | <=65% | 14 | 7% |

### 4.2.1 Dealing with Imbalanced Data

One of the most common preferred approaches to dealing with an imbalanced dataset is to resample the data. Undersampling and oversampling are the two most common ways for this. Oversampling techniques are preferred over undersampling techniques in most cases. The reason for this is that when undersampling the data, a lot of instances from data will be removed, and these removed instances may contain some important information.

#### 4.2.1.1 SMOTE: Synthetic Minority Oversampling Technique

The method of oversampling (SMOTE) is used to create artificial samples of minorities [31]. SMOTE works by selecting the examples in the feature space that are close together, a line between the examples in the feature space is drawing, and drawing a new sample at a point along that line. We applied SMOTE to balance our data.' bad' class is the minority class in our dataset.

### 4.2.2 Feature Selection

We have noticed that Students' performance is impacted by all the features. All features were selected as parameters. So, in the classification algorithms, we used all features in building the predictive model.

### 4.2.3 Data Normalization

All features must be normalized after presuming that they are normally distributed in Bayesian and Parzen-window classifiers. As a result, in the decision-making process, each feature is given equal weight. The mean and standard deviation of the training data are used to normalize the data, assuming the data is Gaussian distributed. To normalize the training data, initially, we computed the mean and the standard deviations to each attribute, or column. Second, we normalized the training data using equation (8).

$$X_{scaled} = \frac{X - mean}{sd} \quad (8)$$

Where mean is the average of x values, sd is the standard deviation of x values, $X_{scaled}$ is the normalized value. The normalization has the benefits that ensure that each feature of the training dataset has a normal distribution with a mean of zero and a standard deviation of one.

### 4.2.4. Experimentation

### A. Pattern Identification

This process includes model training, pattern discovery, testing, and evaluation findings. After dividing the data set into testing and training sets, the prior data set is now ready for use. The classification approaches are used to build the model in the training set. The model is evaluated using a testing set. The results will then be evaluated. We have tested five classification algorithms in order to determine which one of them will work best for the prediction.

### 1-Support Vector Machine (SVM)

SVM is a Supervised Learning algorithm that can be used in classification and regression algorithms, SVM is a discriminative classifier officially defined by a separating hyperplane. In other words, given labeled training data, the algorithm outputs an optimal hyperplane that classifies new examples.

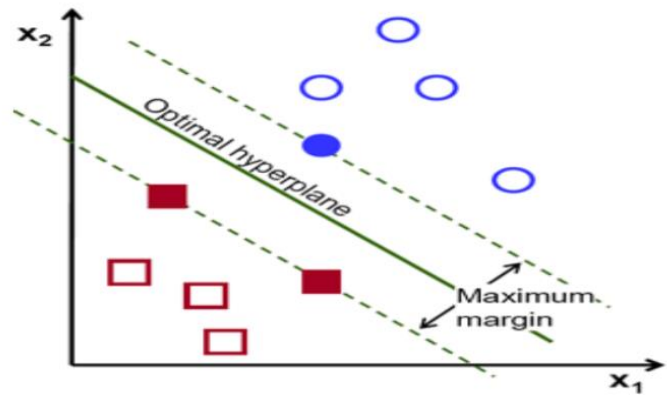

**Figure 6**. SVM [32].

### 2- Logistic Regression (LR)

For many categorization problems, LR is one of the most basic machine learning algorithms and it is a supervised learning algorithm, it utilized to predict the probability of a target variable. A dichotomous is considered the nature of the target or dependent variable, a dichotomous variable has only two possible classes. The nature of binary dependent variable is that its value is single where the data is either 1 (which is for "good" or "yes") or 0 (which is for "bad" or "no"). Mathematically, a logistic regression model predicts P(Y=1) as a function of X.

### 3- Random Forest Classifier (RF)

It is an efficient algorithm that provides more accurate results. Random forest is more efficient because it is the collection of several decision trees. It also prevents the issue of overfitting which is a major issue with decision trees. On the training data, this classifier used the bootstrap sampling method to create a large number of unpruned classification trees. Final predictions are made by using a randomized feature and arithmetic mean of all unpruned classification trees [33]. The following stages describe how Random Forest Algorithm works

- **Stage 1** Start by selecting randomized samples from the dataset.

- **Stage 2** the second step is that a decision tree for each sample will be created. After that, the result of the prediction of each decision tree will be computed.

- **Stage 3** The third stage consists in voting for each expected result.

- **Stage 4** at last, pick the result of the most voted prediction as to the result of the final prediction.

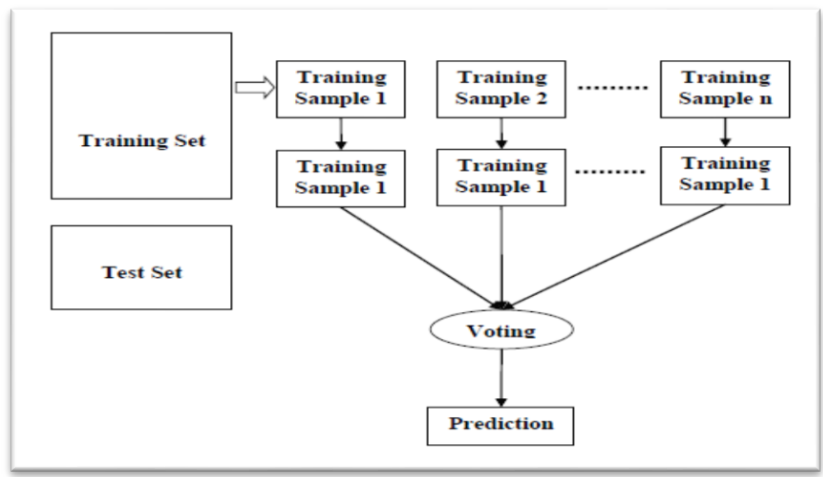

**Figure 7**. Random Forest algorithm.

### 4- Decision Tree (DT)

A Decision tree is a flowchart-like tree structure, in which each internal node designates a test on a characteristic, each branch characterizes the test's result, and each leaf node holds a class label. The

CART algorithm or the C4.5 algorithms can be used to model the decision tree. The items are classified into predefined classes using the decision tree. When it is used to classify data, it is referred to as a classification tree. The decision tree can induce the "If, then" rule to understand the data well and classify them into respective classes correctly [34].

**5- K nearest Neighbors (KNN)**

It is a supervised learning algorithm and one of the simplest machine learning techniques. The KNN algorithm assumes that the new case/data and available cases are similar and places the new case in the category that is most similar to available categories. The KNN algorithm stores all available data and classifies a new data point based on how similar they are to the existing data. This means when new data appears then KNN algorithm can readily classify it into a good group category.

**B. Classifiers Evaluations**

To assess the classification algorithms quality, four different measures were utilized we utilized: accuracy, precision, recall, and F-Measure [35, 36]. measures were derived using Table 4, which shows the classification confusion matrix based on Equations 9, 10, 11, and 12 respectively.

1. Accuracy is the proportion of correct predictions made to the total number of forecasts made.

$$\text{Where, Accuracy} = \frac{TP+TN}{TP+TN+FN+FP} \tag{9}$$

2. Recall is the ratio of positive predictions that are right compared to the total number of positive examples.

$$\text{Where, Recall} = \frac{TP}{TP+FN} \tag{10}$$

3. Precision is the ratio of positive predictions that are right compared to the total number of predicted positives.

$$\text{Where, Precision} = \frac{TP}{TP+FP} \tag{11}$$

4. F-score is a harmonic mean of the model's precision and recall.

$$\text{Where, F-score} = \frac{2*\text{Precision}*\text{Recall}}{\text{Precision}+\text{Recall}} \tag{12}$$

**Table 4.** Confusion Matrix

|  |  | Detected | |
|---|---|---|---|
|  |  | Positive | Negative |
| Actual | Positive | True positive (TP) | False Negative(FN) |
|  | Negative | False Positive (FP) | True Negative (TN) |

### C. Results for the classification algorithms

The data set was separated into ratios of 80% to 20%, having trained data about 80% and testing data about 20%. To predict the performance of each student, five different classification algorithms have been applied to our data. The next Figures and tables show the comparisons between different algorithms.

**Table 5**. The accuracy of different classifiers

| Algorithm | Accuracy |
|---|---|
| RF | 91.7% |
| DT | 84.9% |
| SVM | 78% |
| LR | 75% |
| KNN | 73.9% |

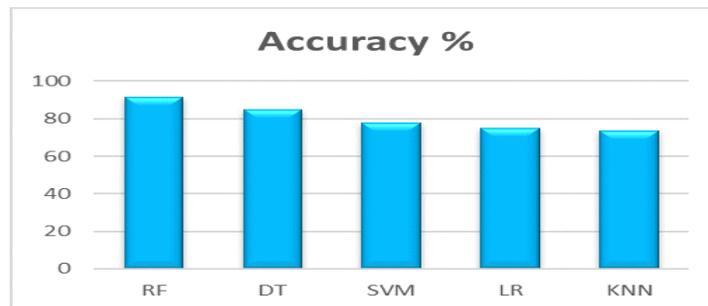

**Figure 8**. The accuracy of different classifiers

Based on the accuracy of the five classifiers, it is evident that the Random Forest classifier beats the others, as shown in Figure 8 and Table 5. It has a 91.7 percent accuracy rate, followed by a decision tree with an accuracy rate of 84.9 %, a support vector machine with a 78 percent accuracy rate, logistic regression with a 75 percent accuracy rate, and lastly K Nearest Neighbors with a 73.9 percent accuracy rate.

**Table 6**. Comparison of different classifiers based on precision, Recall, F-score

| Algorithm | precision | Recall | F-score |
|---|---|---|---|
| RF | 1.0 | 0.85 | 0.92 |
| DT | 0.94 | 0.78 | 0.85 |
| SVM | 1.0 | 0.6097 | 0.75 |
| LR | 0.8965 | 0.6341 | 0.7428 |
| KNN | 0.958 | 0.56 | 0.7076 |

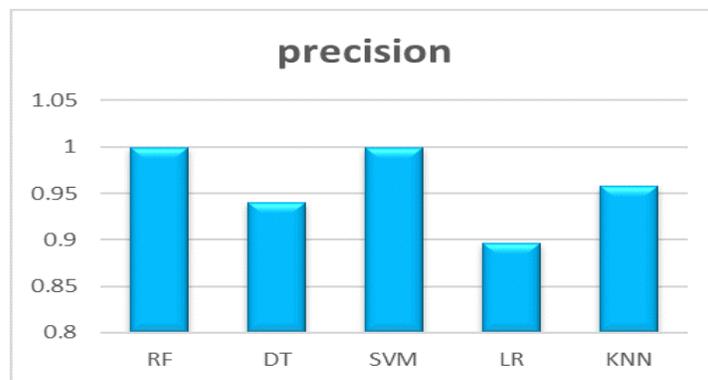

**Figure 9**. Comparison of different classifiers based on precision

Figure 9 shows that, based on the precision of the five classifiers, the Random Forest classifier has the highest precision among them.

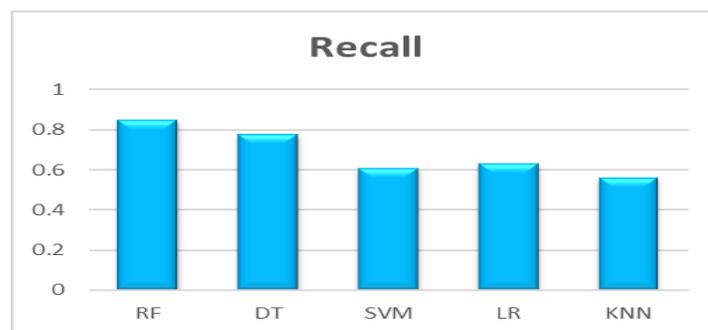

**Figure 10**. Comparison of different classifiers based on Recall.

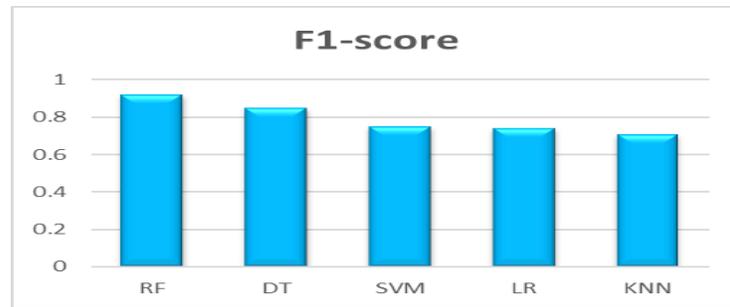

**Figure 11**. Comparison of different classifiers based on F-score.

Figure 10 shows that, based on the recall of the five classifiers, the Random Forest classifier has the highest recall among them.

As shown in Figure 11, we observed that based on the f-score of the five classifiers it can be clear that the Random Forest classifier has the highest f-score among other classifiers.

**D. Discussion for classification Results**

Random Forest classifier, K-nearest neighbors, Support Vector Machine, decision tree, and logistic regression are the five classification methods that were tested. Each one has its own features for classifying the data set. The Random Forest classifier was found to be the best classifier for building our model, outperforming other classifiers in accuracy, precision, recall, and f-score.

## 5. Conclusion

Due to the great spread of the COVID-19 around the world and with the invasion of Technology-enhanced environments, there became an increase in the amount of data in the education sector, notably the data from online eTextbooks, which can be utilized to predict the performance of the student through teaching a particular course. Our study focused on the CS2 course which was taught using an eTextbook. The purpose of our study is to build an early predictive model for students' performance. In this predictive model, Students who are at risk of failure can be detected early, and they can be guided in the right direction for better results in the future. an early Identification of students on the verge of failure will help improve their performance, as well as look for those students who demand particular attention to minimize the failure rate and take required measures to improve their performance. We addressed two problems the prediction of good/bad performance and the prediction of the final exam grade. Both of them aim to early prediction of Students who are at risk of failure.

We used classification to predict the students' performance and regression algorithms to predict students' final exam degrees. In regression analysis, we have evaluated multiple Linear Regression and Random Forest Regression. For classification analysis, we have evaluated Random Forest classifier, Logistic regression, support vector machine, K Nearest Neighbors, and decision tree. We used a correlation coefficient feature selection approach in regression analysis to choose the best features for creating the predictive model. During classification, we noticed that the student's performance is dependent on all features, so all features were selected as parameters in the creation of the classification model.

These different approaches were compared based on their accuracy and error statistics. Based on experimental results, we found that the algorithm with the best result overall in classification was Random Forest Classifier with an accuracy equal to 91.7% while in regression it was Random Forest Regression with $R^2$ equal to 0.977.

**Availability of data and material**

The datasets used and/or analysed during the current study are available from the corresponding author on reasonable request.

**Funding**

There is no fund for this work

**Acknowledgements**

Not applicable

## 6. Future Work

In the future, Experiments can be broadened to incorporate additional distinguishing features in order to acquire more accurate data that can be utilized to improve student learning outcomes. Data mining experiments can be undertaken to gain a broader perspective and produce more valuable results. More datasets will be collected, and various data mining techniques such as clustering and association will be used to compare and analyzed them.